\documentclass[conference]{IEEEtran}
\usepackage[letterpaper, top=2cm, bottom=2.58cm, left=2cm, right=2cm]{geometry}
\IEEEoverridecommandlockouts
\usepackage{cite}
\usepackage{amsmath,amssymb,amsfonts}
\usepackage{algorithmic}
\usepackage{graphicx}
\makeatletter
\newcommand{\removelatexerror}{\let\@latex@error\@gobble}
\makeatother
\usepackage{textcomp}
\usepackage{xcolor}
\usepackage{subcaption}
\usepackage[linewidth=1pt]{mdframed}
\usepackage{graphicx}
\usepackage[export]{adjustbox}
\usepackage[ruled,vlined]{algorithm2e}

\SetAlFnt{\small}
\SetAlCapFnt{\normalsize}
\SetAlCapNameFnt{\normalsize}
\def\BibTeX{{\rm B\kern-.05em{\sc i\kern-.025em b}\kern-.08em
    T\kern-.1667em\lower.7ex\hbox{E}\kern-.125emX}}
\begin{document}

\title{Sim2Real Deep Transfer for Per-Device CFO Calibration}

\author{
\IEEEauthorblockN{Jingze Zheng\IEEEauthorrefmark{1}, Zhiguo Shi\IEEEauthorrefmark{1}, Shibo He\IEEEauthorrefmark{2}, Chaojie Gu\IEEEauthorrefmark{2}}
\IEEEauthorblockA{\IEEEauthorrefmark{1}College of Information Science and Electronic Engineering, Zhejiang University, Hangzhou 310027, China} \IEEEauthorblockA{\IEEEauthorrefmark{2}College of Control Science and Engineering, Zhejiang University, Hangzhou 310027, China} 
\{3190103366, shizg, s18he, gucj\}@zju.edu.cn

\thanks{ This work was supported by the National Natural Science Foundation of China (NSFC) under Grant Nos. (62302439, U23A20296). Chaojie Gu is the corresponding author.}
}

\maketitle

\begin{abstract}
Carrier Frequency Offset (CFO) estimation in Orthogonal Frequency Division Multiplexing (OFDM) systems faces significant performance degradation across heterogeneous software-defined radio (SDR) platforms due to uncalibrated hardware impairments. Existing deep neural network (DNN)-based approaches lack device-level adaptation, limiting their practical deployment. This paper proposes a Sim2Real transfer learning framework for per-device CFO calibration, combining simulation-driven pretraining with lightweight receiver adaptation. A backbone DNN is pre-trained on synthetic OFDM signals incorporating parametric hardware distortions (e.g., phase noise, IQ imbalance), enabling generalized feature learning without costly cross-device data collection. Subsequently, only the regression layers are fine-tuned using $1,000$ real frames per target device, preserving hardware-agnostic knowledge while adapting to device-specific impairments. Experiments across three SDR families (USRP B210, USRP N210, HackRF One) achieve $30\times$ BER reduction compared to conventional CP-based methods under indoor multipath conditions. The framework bridges the simulation-to-reality gap for robust CFO estimation, enabling cost-effective deployment in heterogeneous wireless systems.
\end{abstract}

\section{Introduction}
Orthogonal Frequency Division Multiplexing (OFDM) remains the cornerstone of modern wireless systems, yet its synchronization in software-defined radio (SDR) platforms faces persistent device-specific challenges. Traditional carrier frequency offset (CFO) estimators, whether pilot-aided or blind, often degrade significantly when deployed across heterogeneous hardware. This performance variation stems from uncalibrated RF impairments (e.g., IQ imbalance, oscillator drift and hardware imperfection) that uniquely distort CFO-induced phase rotations in each SDR device.

In common OFDM communication systems, pilot sequences are often used for CFO estimation. The CFO of the signal can be easily obtained by the difference in the pilot symbol between the receiver and the transmitter~\cite{ML}. However, the introduction of a pilot sequence can affect the communication rate. Therefore, the research on OFDM systems that do not rely on pilot sequences has become increasingly popular in recent years~\cite{non-pilot}. Additionally, the blind/non-pilot-assisted schemes are preferred for spectrum efficient and intelligent receiver in a dynamic environment, where several transmitted signal parameters are changed to adapt the time-varying scenarios. This pilot-free design introduces challenges for CFO estimation in OFDM systems, prompting substantial research efforts in this domain~\cite{Blind_CFO_Survey}. Two representative solutions have emerged: the CP-based~\cite{CP-Based} approach leverages the inherent redundancy between the CP and the terminal portion of OFDM symbols to derive frequency offset through cross-correlation analysis of these repeating segments, while the covariance matrix method~\cite{covariance} exploits second-order statistical properties by analyzing structural distortions in the autocorrelation characteristics of received signals caused by frequency offsets.

Recent advances in Deep neural networks (DNNs) offer new possibilities through their inherent capability to learn hardware-specific distortion patterns. However, existing DNN-based CFO estimators predominantly adopt a "one-size-fits-all" approach. To illustrate, \cite{CAD} introduces an CNN-Attention-DNN architecture for CFO estimation and \cite{DNN_dtection} proposes a DNN model for signal detection in OFDM system in the CFO and phase offset (PO) impaired environment at the receiver end. However, these works neglect the critical need for device-level adaptation. Our work bridges this gap through a pragmatic two-stage framework that synergizes simulation-driven pre-training with targeted device fine-tuning~\cite{SDR_fine_tuning}, enabled by three fundamental design principles:

\begin{itemize}
    \item \textbf{Simulation-driven feature learning:} A backbone DNN model is pre-trained on synthetic OFDM signals incorporating parametric hardware impairments (e.g., phase noise, IQ imbalance). This allows the network to learn generalized mappings between raw signal distortions and CFO values, avoiding the prohibitive cost of collecting massive real-world data across devices.

    \item \textbf{Lightweight device adaptation:} For each target SDR device, only the final regression layers are fine-tuned using limited real data (typically $1,000$ frames). Freezing the frontend feature extractor preserves hardware-agnostic knowledge while adapting to device-specific RF characteristics.
    
     \item \textbf{Receiver-centric fingerprinting:} Unlike transmitter-oriented pre-distortion algorithms that require closed-loop calibration, our design specifically targets receiver-side hardware fingerprints. There is no doubt that if we learn the fingerprints of transceiver pair we can improve the performance of CFO. However, this design lacks generalizability as it only functions when a specific device serves as transmitter and another specific SDR serves as receiver. According to our experiment, the receiver-specific design performs well when various SDR devices serve as transmitter.
    
\end{itemize}

This approach demonstrates four key advantages over state-of-the-art: 1) Eliminates pairwise transceiver calibration, 2) Maintains backward compatibility through additive DNN integration, 3) Requires only 5-7\% additional parameters per device, and 4) Enables lifelong service through continuous receiver-specific adaptation. Experimental validation demonstrates strong correlation between simulation pre-training accuracy and real-world performance across $3$ SDR families including USRP B210, USRP N210 and HackRF One with $2$ devices for each kind of SDR platform, confirming the framework's device-agnostic viability.

The remainder of this paper is organized as follows: Section \ref{sec:System Model} analyzes device-specific CFO distortion mechanisms. Section \ref{sec:method} details the DNN architecture and transfer learning strategy. Section \ref{Performance Benchmarks} presents performance benchmarks, followed by evaluations in Section \ref{sec:fine_tuning} and Section \ref{sec:conclision} as conclusion.
\section{System Model}
\label{sec:System Model}
\begin{figure*}[t]
\centering
\includegraphics[width=0.87\linewidth]{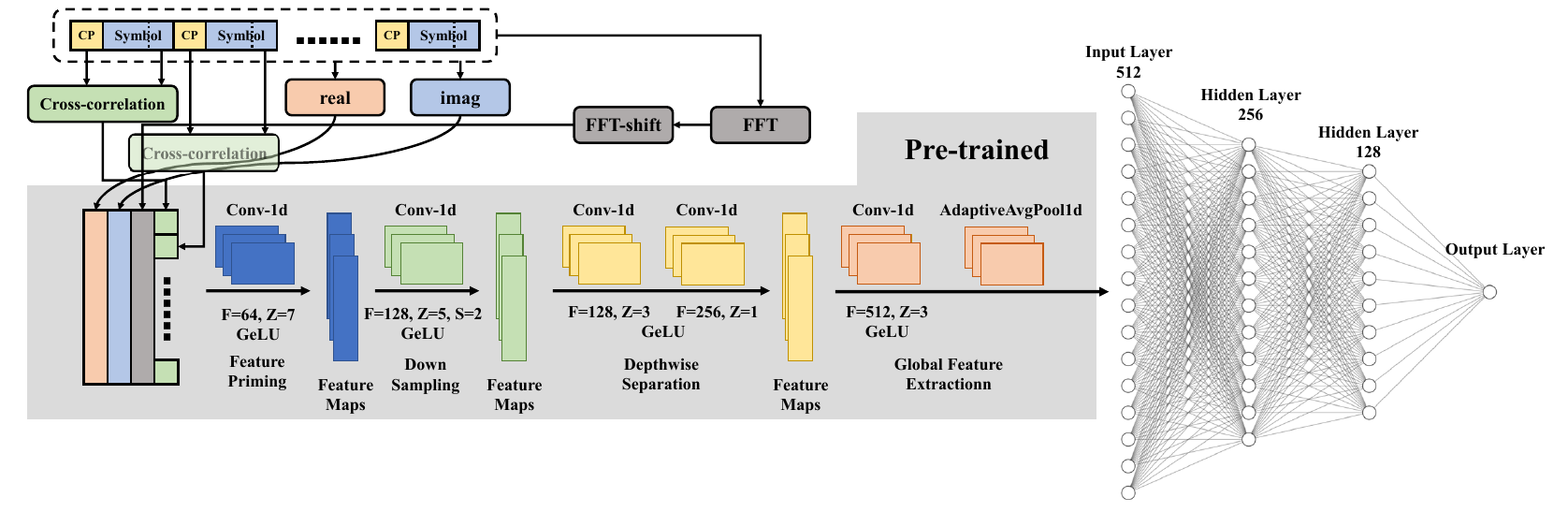}
\caption{DNN architecture for CFO estimation.}
\label{fig:arch}
\end{figure*}
\subsection{OFDM Signal Structure}
Consider an OFDM system with symbol length $K$ and cyclic prefix length $G$. Let $\mathbf{s} \in \mathbb{C}^{K}$ denote the time-domain symbol vector before CP insertion. The complete OFDM symbol with CP can be expressed as:

\begin{equation}
    \mathbf{x} = \left[ \mathbf{s}_{K-G+1:K}^T \ \mathbf{s}^T \right]^T \in \mathbb{C}^{K+G}
\end{equation}

where $\mathbf{s}_{K-G+1:K}$ represents the last $G$ samples of $\mathbf{s}$. The time-domain signal experiences CFO-induced phase rotation given by:

\begin{equation}
    \theta = \frac{\Delta f}{f_s} \cdot \frac{K}{2\pi}
\end{equation}

where $\Delta f$ denotes the frequency offset and $f_s$ the sampling rate.

\subsection{Received Signal Model}
The received signal $\mathbf{r} \in \mathbb{C}^{K+G}$ after transmission through multipath channel $\mathbf{h}$ with $L$ taps can be written as:

\begin{IEEEeqnarray}{rCl}
    \mathbf{r} & = & e^{j2\pi\theta\mathbf{\Gamma}}\mathbf{H}\mathbf{x} + \mathbf{n}, \IEEEyesnumber \\
    \mathbf{\Gamma} & = & \text{diag}(0,1,\ldots,K+G-1)^T, \IEEEyessubnumber
\end{IEEEeqnarray}

where $\mathbf{H} \in \mathbb{C}^{(K+G)\times(K+G)}$ is the circulant channel matrix, $\mathbf{n} \sim \mathcal{CN}(0,\sigma_n^2\mathbf{I})$ represents additive white Gaussian noise, and the diagonal matrix $e^{j2\pi\theta\mathbf{\Gamma}}$ captures cumulative CFO effects.

\subsection{CP-Based Feature Extraction}
The proposed method exploits the inherent CP structure through three key operations:

\begin{enumerate}
    \item \textbf{Conjugate Multiplication}: For received CP segment $\mathbf{r}_c$ and corresponding symbol tail $\mathbf{r}_t$:
    \begin{equation}
        \mathbf{m} = \mathbf{r}_c \odot \mathbf{r}_t^*.
    \end{equation}
    
    \item \textbf{Phase Sequence Derivation}:
    \begin{equation}
        \mathbf{\Phi} = \arg(\mathbf{m}) = 2\pi\theta\mathbf{I}_G + \boldsymbol{\omega},
    \end{equation}
    where $\boldsymbol{\omega}$ contains combined noise and channel effects (e.g., multi-path effect).
    
\end{enumerate}

\subsection{Problem Formulation}
\label{subsec:problem}

\paragraph{Conventional CP-Based Estimation} 
The classical approach exploits CP's phase coherence through maximum likelihood (ML) estimation:

\begin{IEEEeqnarray}{rCl}
    \hat{\theta}_{\text{ML}} &=& \frac{1}{2\pi}\arg\left(\mathbf{m}\right) \IEEEyesnumber \\
    &=& \theta + \underbrace{\frac{1}{2\pi}\arg\left(1 + \frac{\sum \epsilon_k}{\sum |r_t[k]|^2}\right)}_{\text{Noise perturbation term}}. \IEEEyessubnumber
    \label{conventional CP-based}
\end{IEEEeqnarray}

where $\epsilon_k$ represents combined device, noise and channel distortion. This estimator suffers from two fundamental limitations:
\begin{enumerate}
    \item \textbf{Nonlinear Squaring Loss}: The argument operation $\arg(\cdot)$ amplifies phase noise at low SNR,
    \item \textbf{Single-Point Estimation}: Utilizes only the summation result, discarding per-sample phase information.
\end{enumerate}
This simplification inherently neglects the noise perturbation term under the assumption of zero-mean additive noise characteristics, while deliberately omitting the non-ideal hardware-induced distortions from SDR frontends and residual channel artifacts. Such abstraction enables tractable theoretical analysis but necessitates neural network compensation in practical implementations to address unmodeled error components.

\paragraph{Deep Learning Reformulation} 
For deep learning consideration, the CFO estimation problem is formulated as a nonlinear regression task:

\begin{equation}
    \hat{\theta} = \mathcal{F}_\Theta(\mathbf{\Phi};\mathbf{W}),
\end{equation}
where $\mathcal{F}_\Theta$ represents the deep neural network parameterized by weights $\mathbf{W}$, trained to minimize:

\begin{equation}
    \mathcal{L}(\Theta) = \mathbb{E}\left[ (\theta - \hat{\theta})^2 \right] + \lambda\|\mathbf{W}\|_F^2,
\end{equation}

The key challenge lies in learning robust mappings from noise-corrupted phase sequences $\mathbf{\Phi}$ under varying SNR conditions and hardware impairments.

\section{DNN Architecture and Transfer Learning Strategy}
\label{sec:method}

\subsection{Network Architecture}
The proposed CFO estimation network, depicted in Fig.~\ref{fig:arch}, employs a hybrid convolutional-fully connected structure optimized for hardware-robust feature extraction:

\begin{itemize}
    \item \textbf{Convolutional Frontend}: Processes the phase information of raw I/Q samples through three stages:
    \begin{enumerate}
        \item \textit{Feature Priming}: 64-channel 1D convolution (kernel=7) with batch normalization and GELU activation
        \item \textit{Downsampling}: Strided convolution (kernel=5, stride=2) increases receptive field while preserving temporal resolution
        \item \textit{Depthwise Separation}: Factorized into channel-wise spatial convolution (kernel=3) followed by pointwise fusion
    \end{enumerate}
    
    \item \textbf{Global Pooling}: Adaptive average pooling compresses spatial dimensions while retaining device-specific signatures:
    \begin{equation}
        \mathbf{f}_g = \frac{1}{T}\sum_{t=1}^T \mathbf{F}[:,t],
    \end{equation}
    where $\mathbf{F} \in \mathbb{R}^{512 \times T}$ is the final convolutional feature map.
    
    \item \textbf{Regression Head}: Three fully-connected layers with progressive dimension reduction ($512\to256\to128\to1$):
    \begin{equation}
        \hat{\theta} = \mathbf{W}_3\sigma(\mathbf{W}_2\sigma(\mathbf{W}_1\mathbf{f}_g + \mathbf{b}_1) + \mathbf{b}_2) + \mathbf{b}_3,
    \end{equation}
    where $\sigma$ alternates between GELU and SiLU activations. The complete training procedure is summarized in Algorithm~\ref{alg:training}.
\end{itemize}

\subsection{Transfer Learning Strategy} 
The adaptation process employs two-phase optimization:

\begin{itemize}
    \item \textbf{Pre-training}: On synthetic dataset $\mathcal{D}_{\text{sim}}$ with multi-device impairments:
    \begin{equation}
        \mathcal{L}_{\text{pre}} = \frac{1}{|\mathcal{D}_{\text{sim}}|}\sum_{(\mathbf{\Phi}_i,\theta_i)\in\mathcal{D}_{\text{sim}}} \|\theta_i - \hat{\theta}_i\|_2^2 + \lambda\|\mathbf{W}\|_F^2.
    \end{equation}
    
    \item \textbf{Fine-tuning}: Freeze convolutional layers $\mathbf{W}_{\text{conv}}$, optimize only FC head using $\mathcal{D}_{\text{real}}$:
    \begin{equation}
        \mathbf{W}_{\text{FC}}^* = \arg\min_{\mathbf{W}_{\text{FC}}} \mathcal{L}_{\text{FC}}.
    \end{equation}
    However, the fundamental distinction between pre-training and fine-tuning phases stems from the \textit{label uncertainty} in real-device adaptation. Unlike the simulation phase where ground-truth CFO values $\theta_{\text{sim}}$ are precisely known, the SDR-transmitted signals introduce unobservable hardware perturbations $\Delta\theta_{\text{hw}}$ that prevent direct supervised learning. Therefore, we re-design the loss $\mathcal{L}_{\text{FC}}$ as:
    \begin{equation}
        \mathcal{L}_{\text{FC}} = \frac{1}{|\mathcal{D}_{\text{FC}}|}\sum_{\mathcal{D}_{\text{i}}\in\mathcal{D}_{\text{FC}}}\|f_D(\mathcal{D}_{\text{i}})-S_t\|^2,
    \end{equation}   
    where $f_D$ represents the OFDM demodulation function, which is determined by OFDM configurations (e.g., symbol length, CP length and subcarrier modulation) and $S_t$ represents the symbols transmitted.  
\end{itemize}

\subsection{Implementation Details}
\begin{itemize}
    \item \textbf{Initialization}: Convolutional weights follow He normal with fan-out mode, biases initialized to $0.1$;
    \item \textbf{Regularization}: Varied dropout rates ($0.25-0.4$) prevent co-adaptation to simulated distortions;
    \item \textbf{Activation}: GELU in early layers enhances gradient flow, SiLU in final FC layer suppresses phase wrapping.
\end{itemize}
\begin{figure}[h]
	\renewcommand{\algorithmicrequire}{\textbf{Input:}}
	\renewcommand{\algorithmicensure}{\textbf{Output:}}
	\removelatexerror
	\begin{algorithm}[H]
    \label{alg:training}
		\caption{DNN Training with L2 Regularization for resource-constrained Deployment}
		\begin{algorithmic}[1]
			\REQUIRE Train set $\mathcal{D}_{\text{train}}$, Validation set $\mathcal{D}_{\text{val}}$, Regularization coefficient $\lambda = 10^{-4}$
			\ENSURE Optimal model parameters $\Theta^*$  
			\STATE {Initialize CNN $\mathcal{F}_\Theta$: He initialization for convolutional layers, bias initialized to $0.1$}
            \STATE {Configure optimizer: AdamW (learning rate $\eta= 0.001$), ReduceLROnPlateau scheduler, Gradient Scaler, Gradient clipping threshold $\|\nabla\|_2 \leq 1.0$}
            \FOR{epoch $e = 1$ \textbf{to} $N_{\text{epochs}}$}
                    \FOR{each mini-batch $\mathcal{B} \subset \mathcal{D}_{\text{train}}$}
                       \STATE Standardize input features $\mathbf{X}_b$
                       \STATE Compute predicted CFO $\hat{\theta}_b = \mathcal{F}_\Theta(\mathbf{X}_b)$
                       \STATE Calculate loss $\mathcal{L} = \text{MSE}(\theta_b, \hat{\theta}_b) + \lambda\|\Theta\|_2^2$
                    \ENDFOR

            \STATE{Update learning rate scheduler, save best parameters $\Theta^*$ if validation loss improves}
            \ENDFOR
 
		\end{algorithmic}
	\end{algorithm}
\end{figure}

\section{Performance Benchmarks}
\label{Performance Benchmarks}

To comprehensively evaluate the proposed DNN-based frequency offset estimation algorithm, we conducted extensive simulations under varying SNR conditions. The training dataset encompassed SNR levels ranging from $0$ dB to $30$ dB, while specifically generated test sets at $0$, $3$, $6$, $9$, and $12$ dB SNR levels were used for performance validation.

\begin{figure}[t]
    \centering
    \includegraphics[width=0.85\linewidth]{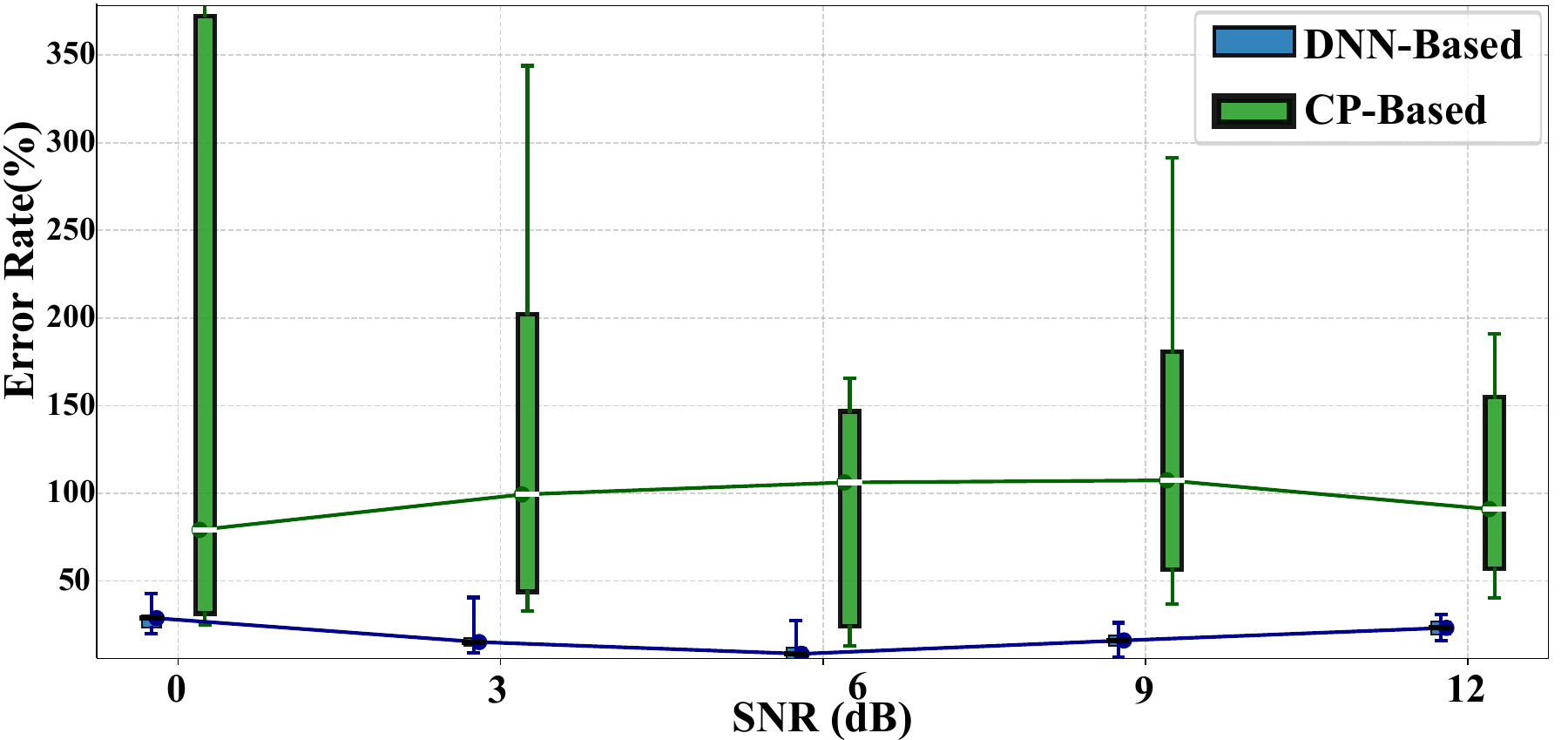}
    \caption{Comparison of estimation error distributions between conventional CP-based method and proposed DNN approach across different SNR levels.}
    \label{fig_sim}
\end{figure}

The comparative analysis in Fig.~\ref{fig_sim} reveals fundamental distinctions in algorithmic robustness between the competing approaches. The boxplots demonstrate superior error concentration of our method, particularly in low-SNR conditions. The proposed DNN estimator demonstrates remarkable consistency across all simulated SNR conditions, maintaining tightly clustered error distributions that contrast sharply with the erratic performance patterns of conventional CP-based estimation. Notably in low-SNR regimes, the neural network approach exhibits inherent noise immunity, preserving stable error characteristics while the traditional method succumbs to severe performance degradation manifested through outlier proliferation and quartile expansion. This divergence intensifies with decreasing SNR levels, ultimately leading to complete operational failure of the CP-based technique in sub-6 dB environments. The interquartile range analysis further confirms the DNN's superior estimation stability, showing consistently compact error distributions that translate to reliable real-world applicability. 

To quantify estimator consistency, we calculate the error variance and compare with the Cramér-Rao Lower Bound (CRLB)~\cite{CRLB}. This theoretical bound characterizes the minimum achievable variance for any unbiased estimator through the Fisher information matrix:
\begin{equation}
    \mathrm{CRLB}=\frac{1}{(2\pi)^2\cdot\mathrm{SNR}\cdot\sum_{n=0}^{N-1}n^2|x[n]|^2}.
\end{equation}

The CRLB's inverse proportionality to SNR ($\text{CRLB} \propto \frac{1}{\text{SNR}}$) establishes two fundamental principles:
\begin{itemize}
    \item Theoretical optimality limit for linear unbiased estimators,
    \item Exponential sensitivity to observation length $N$.
\end{itemize}

\begin{table}[!t]
\caption{Performance Comparison: Estimation Variance vs. CRLB (Hz\textsuperscript{2})}
\label{tab1}
\centering
\begin{tabular}{c|c|c|c}
\hline
SNR (dB) & DNN & CP-based & CRLB \\ 
\hline
0 & 5455.0 & 154897.1 & 68.5 \\
3 & 833.5 & 141586.4 & 35.3 \\
6 & 97.8 & 119271.8 & 17.2 \\
9 & 1669.7 & 96231.0 & 8.8 \\
12 & 4610.7 & 67862.2 & 4.4 \\
\hline
\end{tabular}
\end{table}

The experimental results in Table~\ref{tab1} demonstrate that the DNN-based estimation method closely approaches the theoretical CRLB at $6\,\text{dB}$ SNR. More importantly, while exhibiting minor performance fluctuations in higher SNR regimes ($9-12\,\text{dB}$), the proposed approach maintains substantial performance enhancement over conventional CP-based methods across all tested conditions. Additionally, these fluctuations are acceptable as the fine-tuning step improves the performance of the model under high-SNR situations.

These results validate that the data-driven approach successfully overcomes the limitations of CP-based estimation, particularly in challenging low-SNR scenarios.

\section{Fine-Tuning and Evaluation}
\label{sec:fine_tuning}
\begin{figure*}[h]
  \centering
  \begin{subfigure}{.23\textwidth}
    \centering
    \includegraphics[width=\linewidth,height=0.75\linewidth,frame]{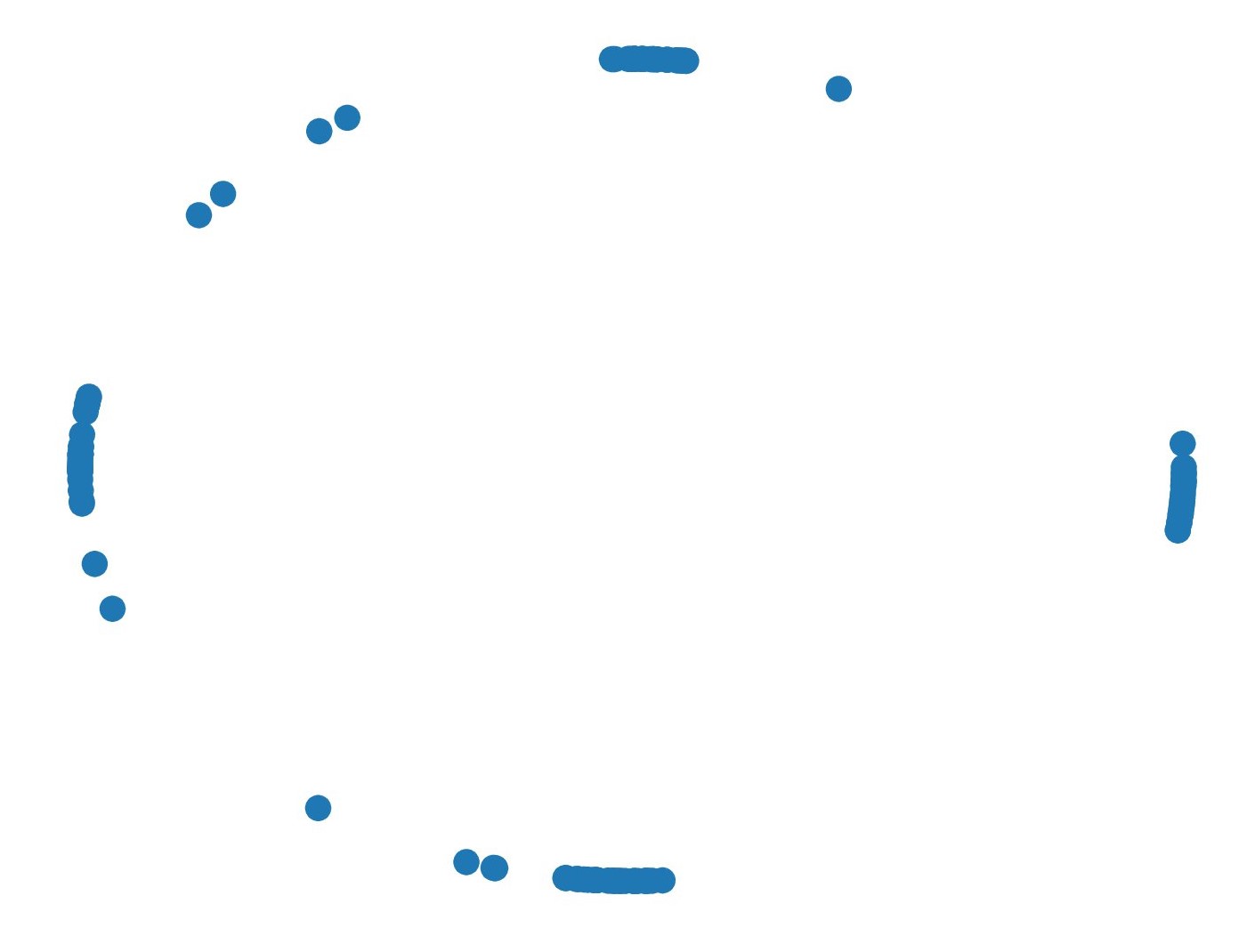}
    \caption{\label{USRP N210 CP-based}USRP N210 RX with CP-based estimation.}
  \end{subfigure}
  \begin{subfigure}{.23\textwidth}
    \centering
    \includegraphics[width=\linewidth,height=0.75\linewidth,frame]{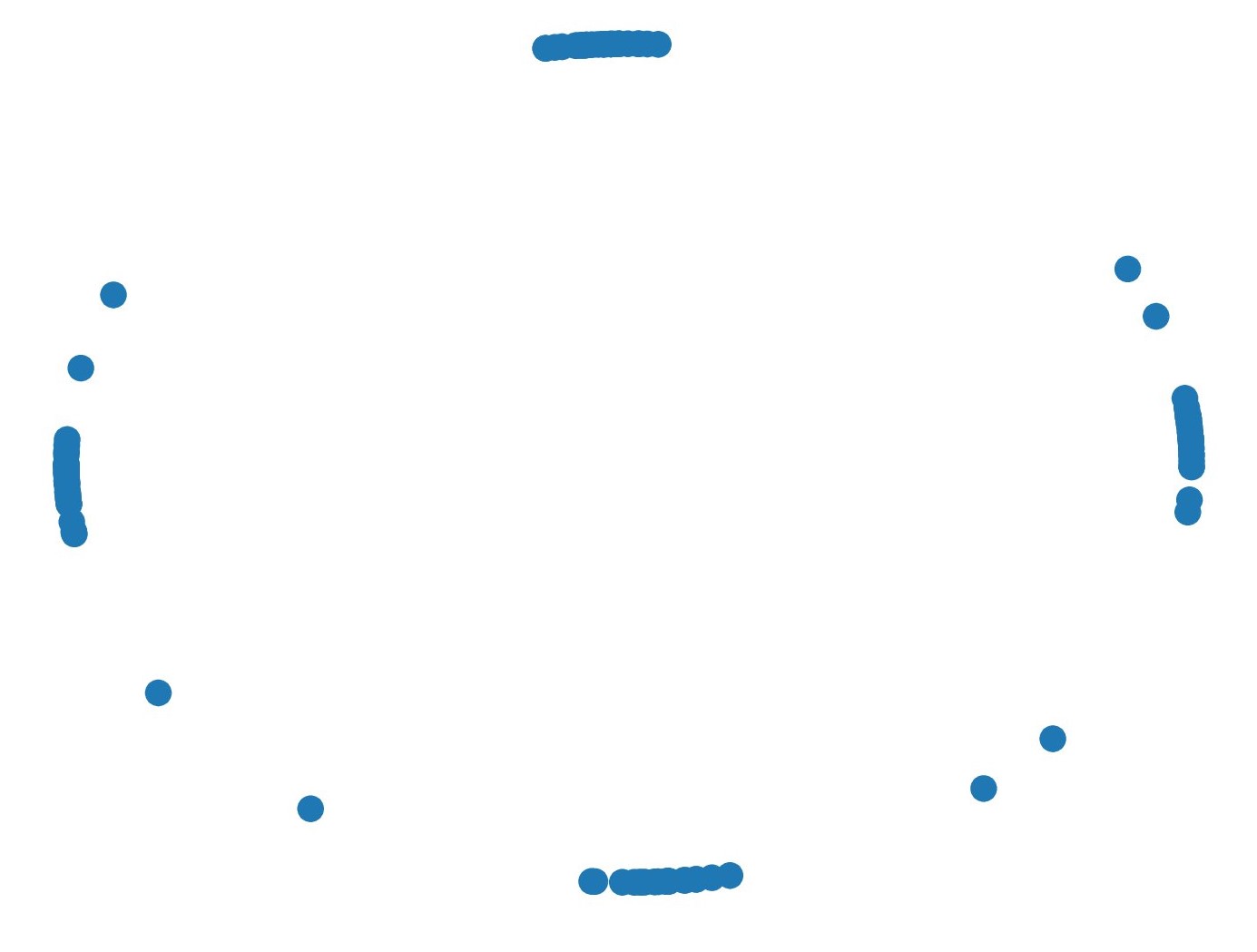}
    \caption{\label{USRP N210 pre-trained network}USRP N210 RX with pre-trained network estimation.}
  \end{subfigure}
  \begin{subfigure}{.23\textwidth}
    \centering
    \includegraphics[width=\linewidth,height=0.75\linewidth,frame]{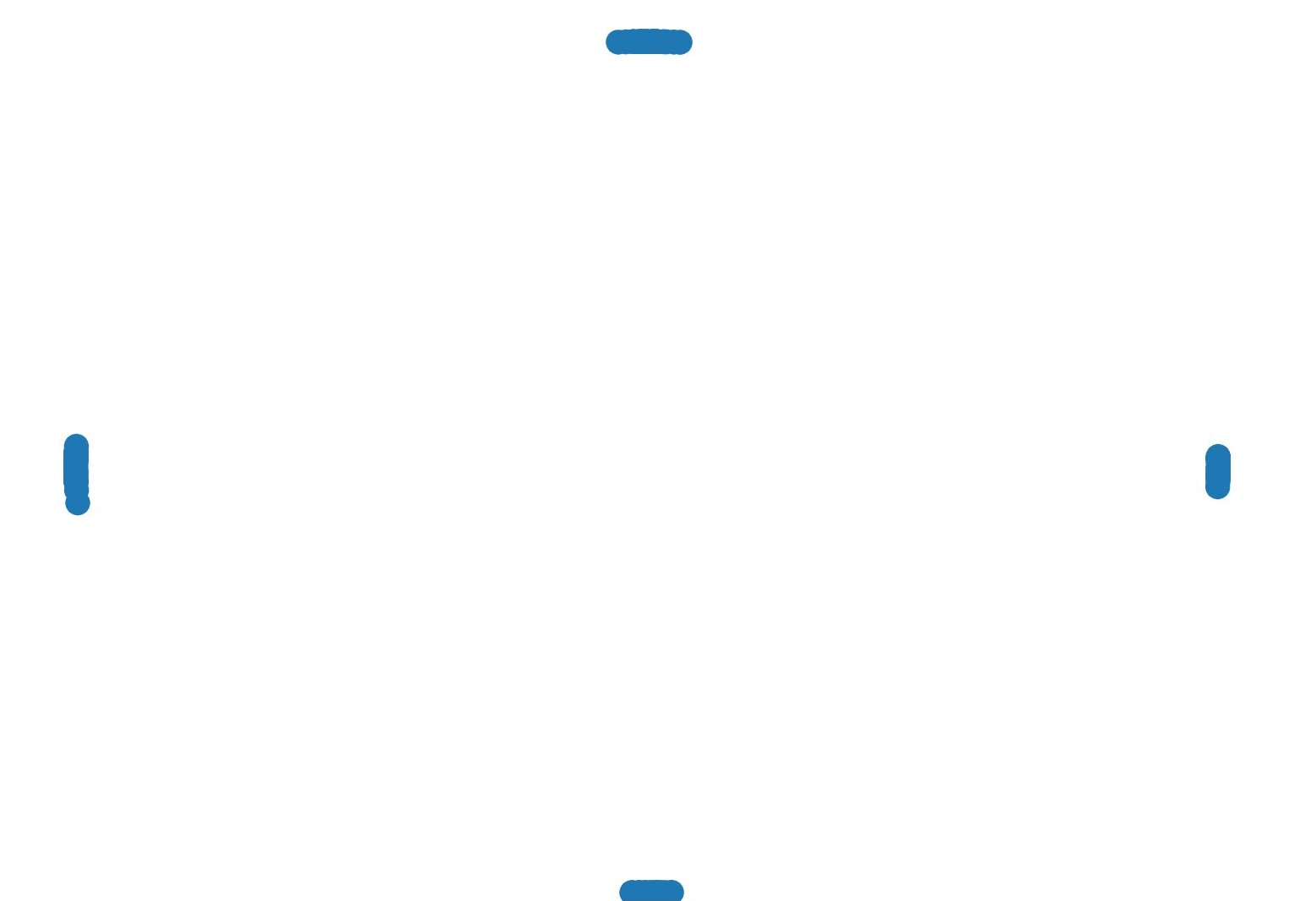}
    \caption{\label{USRP N210 fine-tuned network}USRP N210 RX with fine-tuned network estimation.}
  \end{subfigure}\\
    \begin{subfigure}{.23\textwidth}
    \centering
    \includegraphics[width=\linewidth,height=0.75\linewidth,frame]{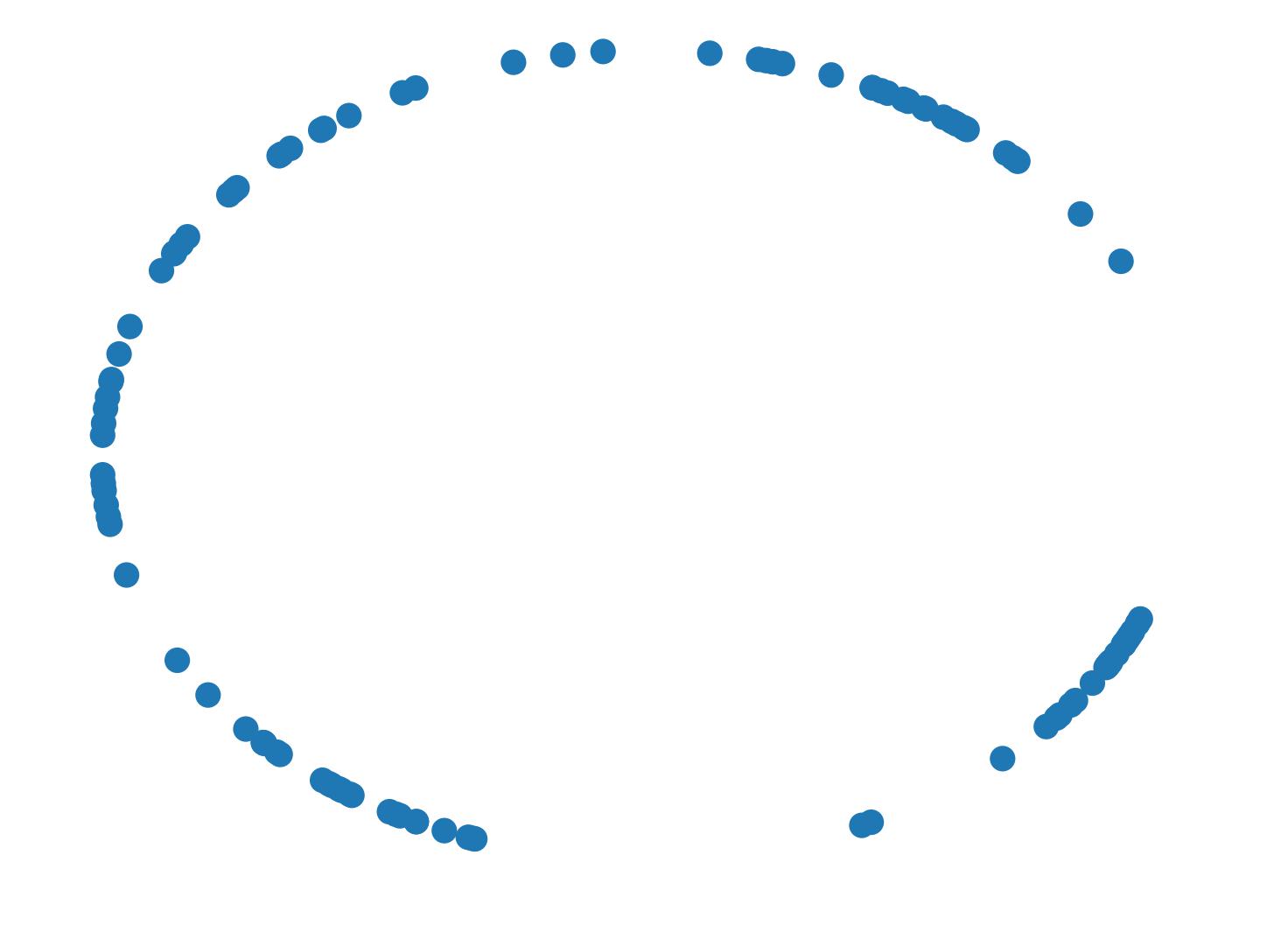}
    \caption{\label{USRP B210 CP-based}USRP B210 RX with CP-based estimation.}
  \end{subfigure}
  \begin{subfigure}{.23\textwidth}
    \centering
    \includegraphics[width=\linewidth,height=0.75\linewidth,frame]{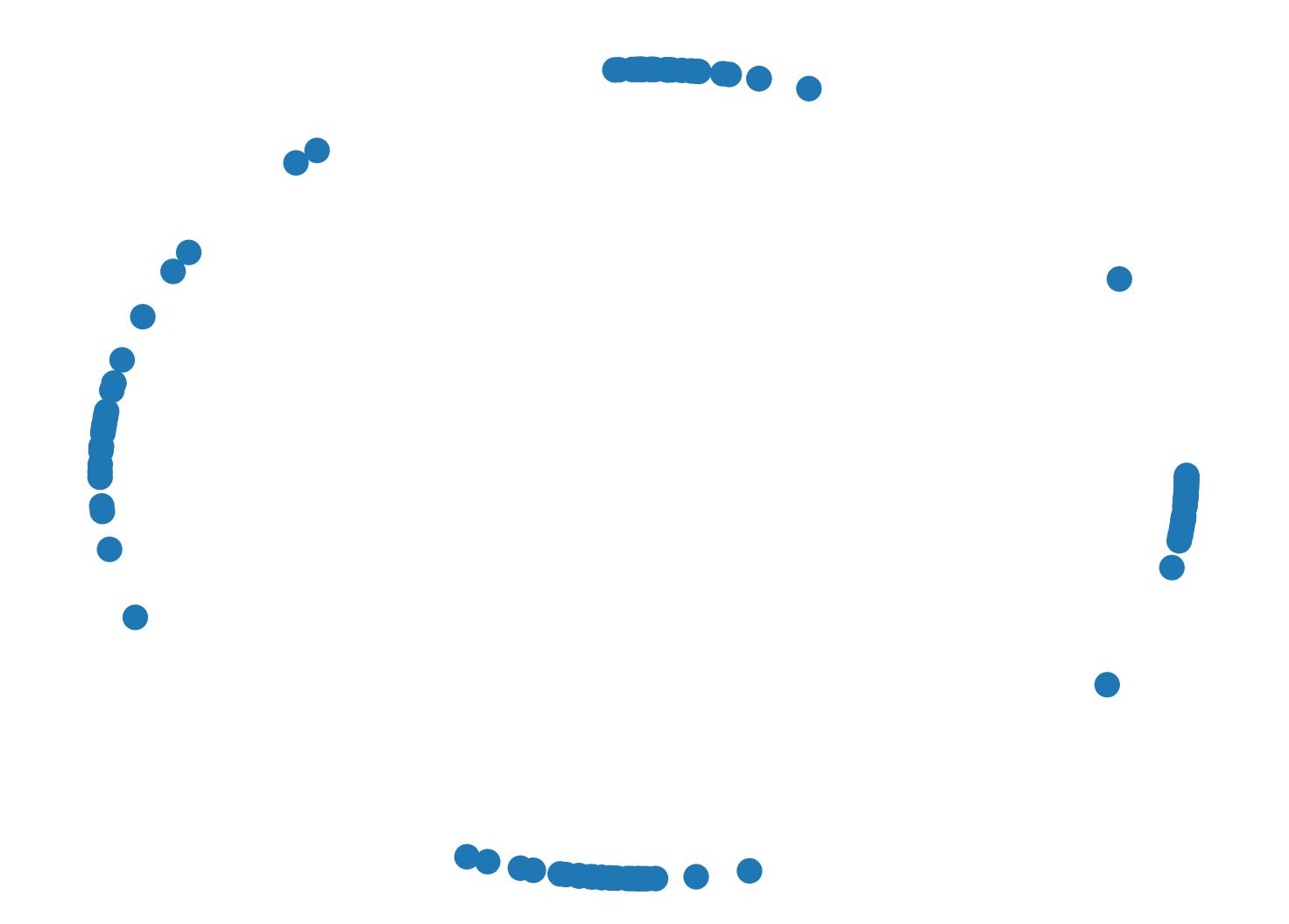}
    \caption{\label{USRP B210 pre-trained network}USRP B210 RX with pre-trained network estimation.}
  \end{subfigure}
  \begin{subfigure}{.23\textwidth}
    \centering
    \includegraphics[width=\linewidth,height=0.75\linewidth,frame]{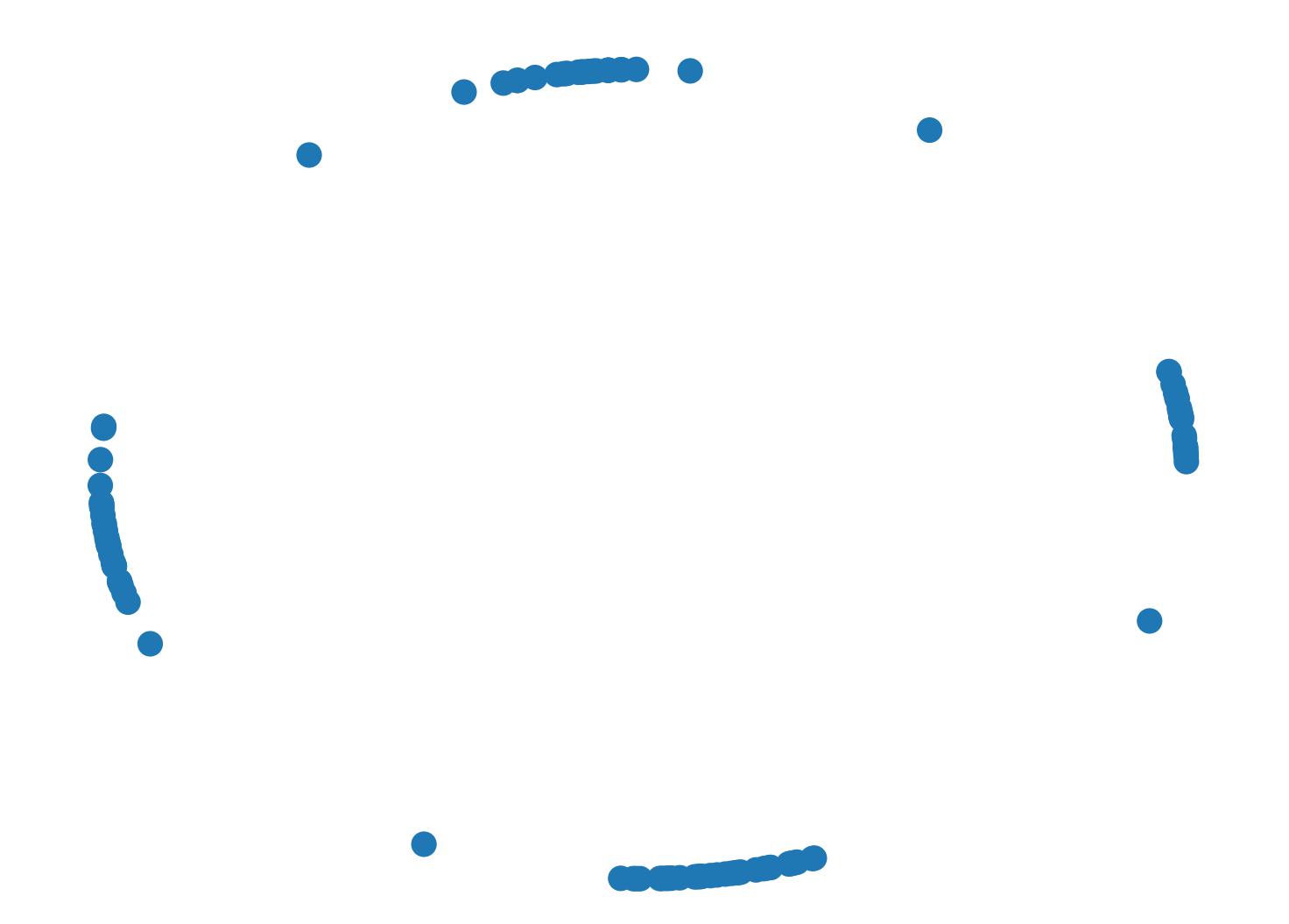}
    \caption{\label{USRP B210 fine-tuned network}USRP B210 RX with fine-tuned network estimation.}
  \end{subfigure}\\
      \begin{subfigure}{.23\textwidth}
    \centering
    \includegraphics[width=\linewidth,height=0.75\linewidth,frame]{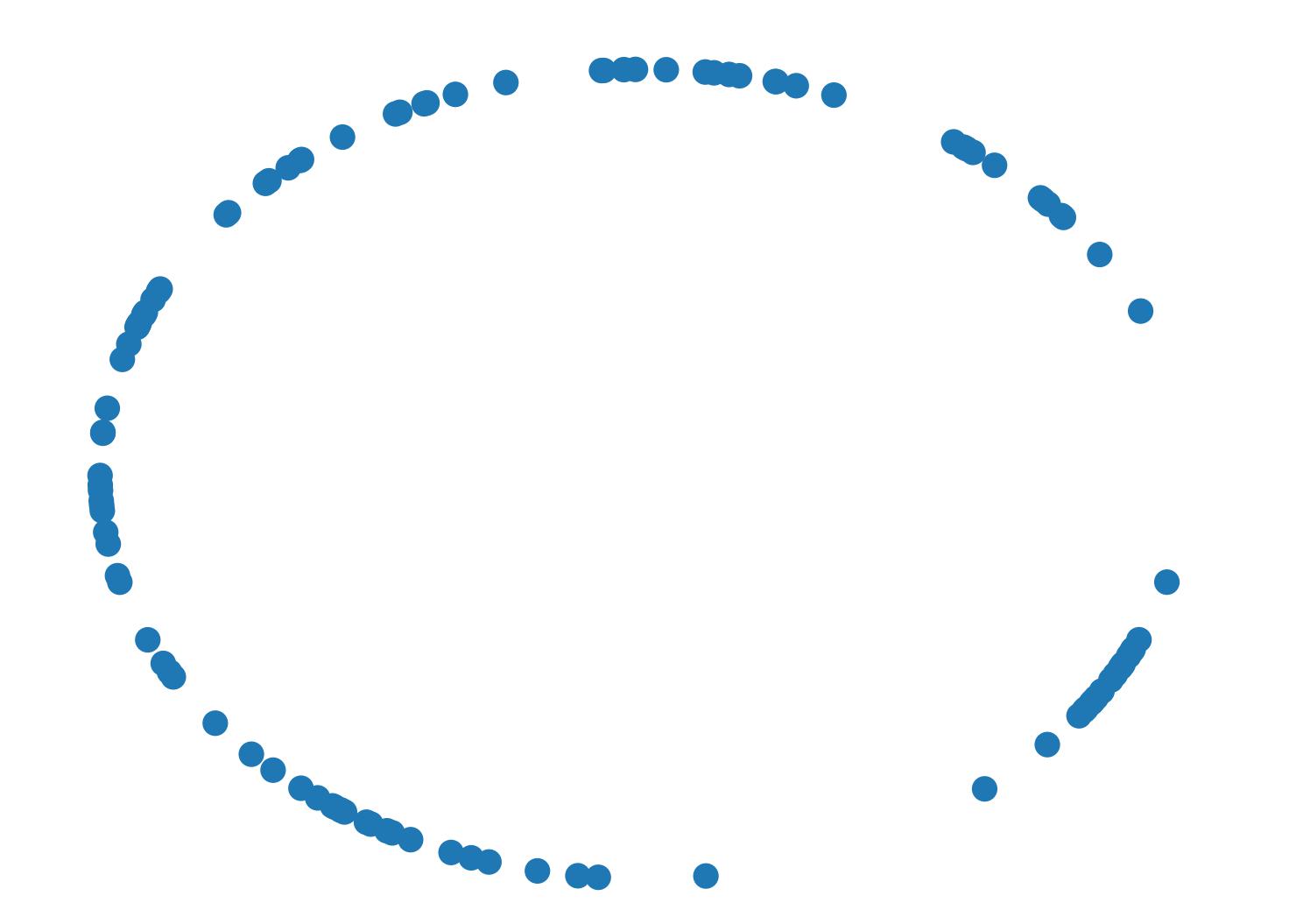}
    \caption{\label{HackRF CP-based}HackRF RX with CP-based estimation.}
  \end{subfigure}
  \begin{subfigure}{.23\textwidth}
    \centering
    \includegraphics[width=\linewidth,height=0.75\linewidth,frame]{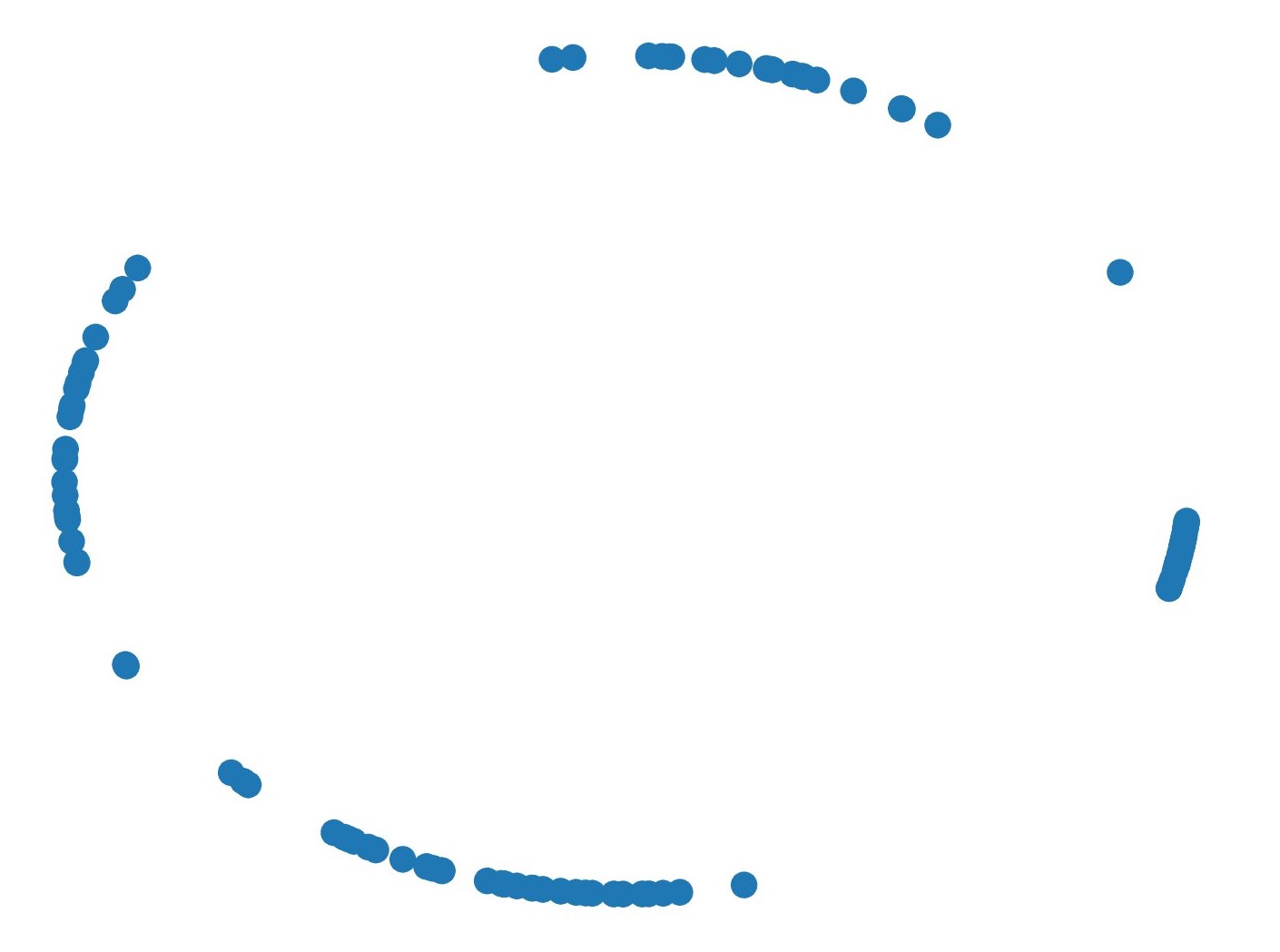}
    \caption{\label{HackRF pre-trained network}HackRF RX with pre-trained network estimation.}
  \end{subfigure}
  \begin{subfigure}{.23\textwidth}
    \centering
    \includegraphics[width=\linewidth,height=0.75\linewidth,frame]{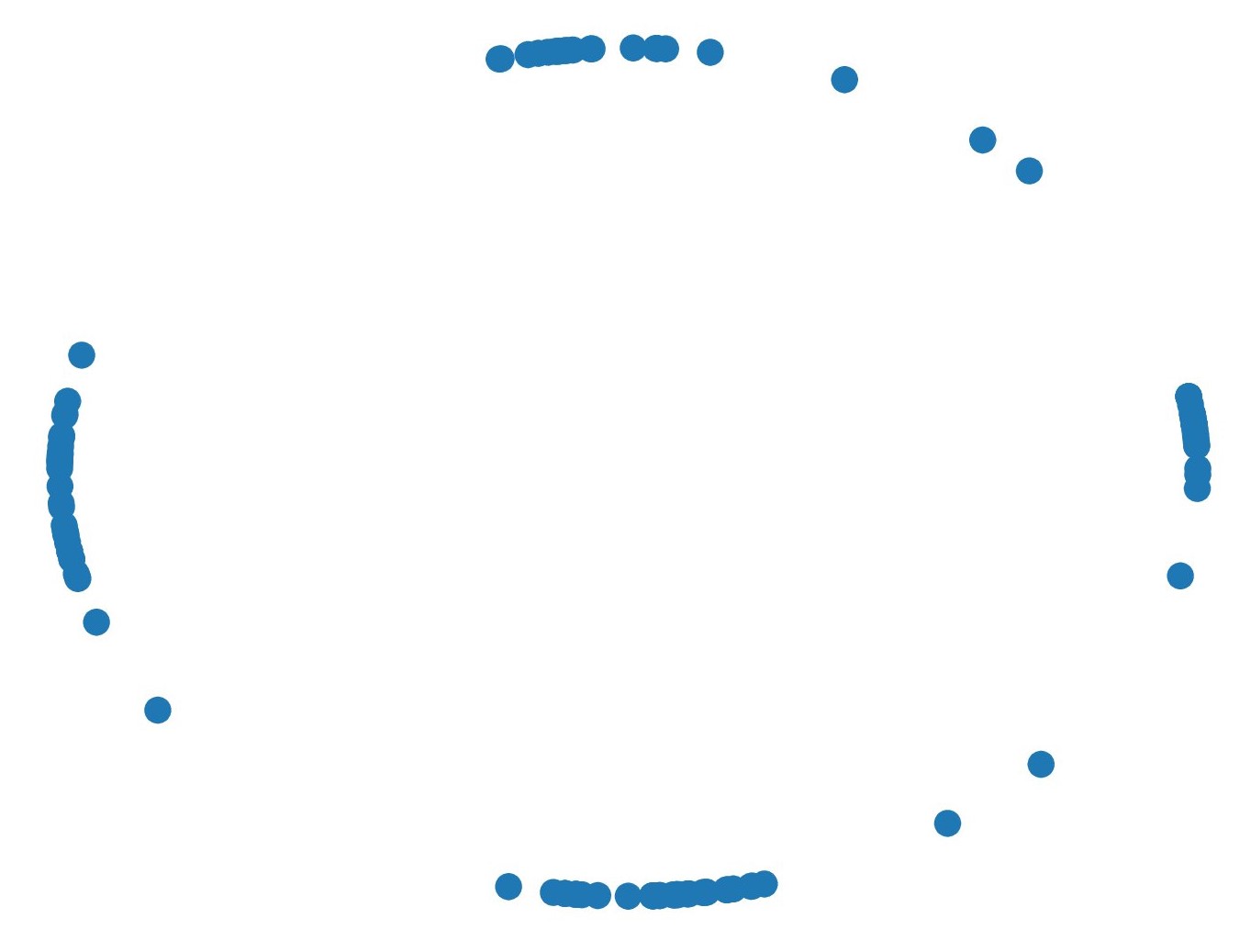}
    \caption{\label{HackRF fine-tuned network}HackRF RX with fine-tuned network estimation.}
  \end{subfigure}
  \caption{Demodulation result after CFO compensation.}
  \label{fig:Rx_constellation}
\end{figure*}
\subsection{Device-Specific Fine-Tuning Protocol}
\begin{figure}[h]
  \centering
  \begin{subfigure}{.15\textwidth}
    \centering
    \includegraphics[width=\linewidth]{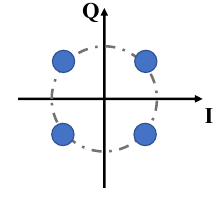}
    \caption{\label{fig:QPSK}QPSK Symbols transmitted.}
  \end{subfigure}
  \begin{subfigure}{.15\textwidth}
    \centering
    \includegraphics[width=\linewidth]{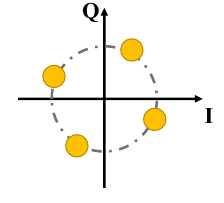}
    \caption{\label{fig:QPSK_Rotation}Symbols at receiver.}
  \end{subfigure}
  \begin{subfigure}{.15\textwidth}
    \centering
    \includegraphics[width=\linewidth]{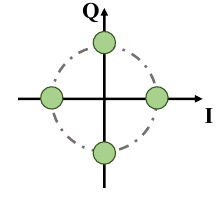}
    \caption{\label{fig:DQPSK}Differential symbols.}
  \end{subfigure}
  \caption{The transmission of DQPSK Symbols.}
  \label{fig:DQPSK_transmission}
\end{figure}

\begin{figure}[h]
  \centering
  \begin{subfigure}{.15\textwidth}
    \centering
    \includegraphics[width=\linewidth]{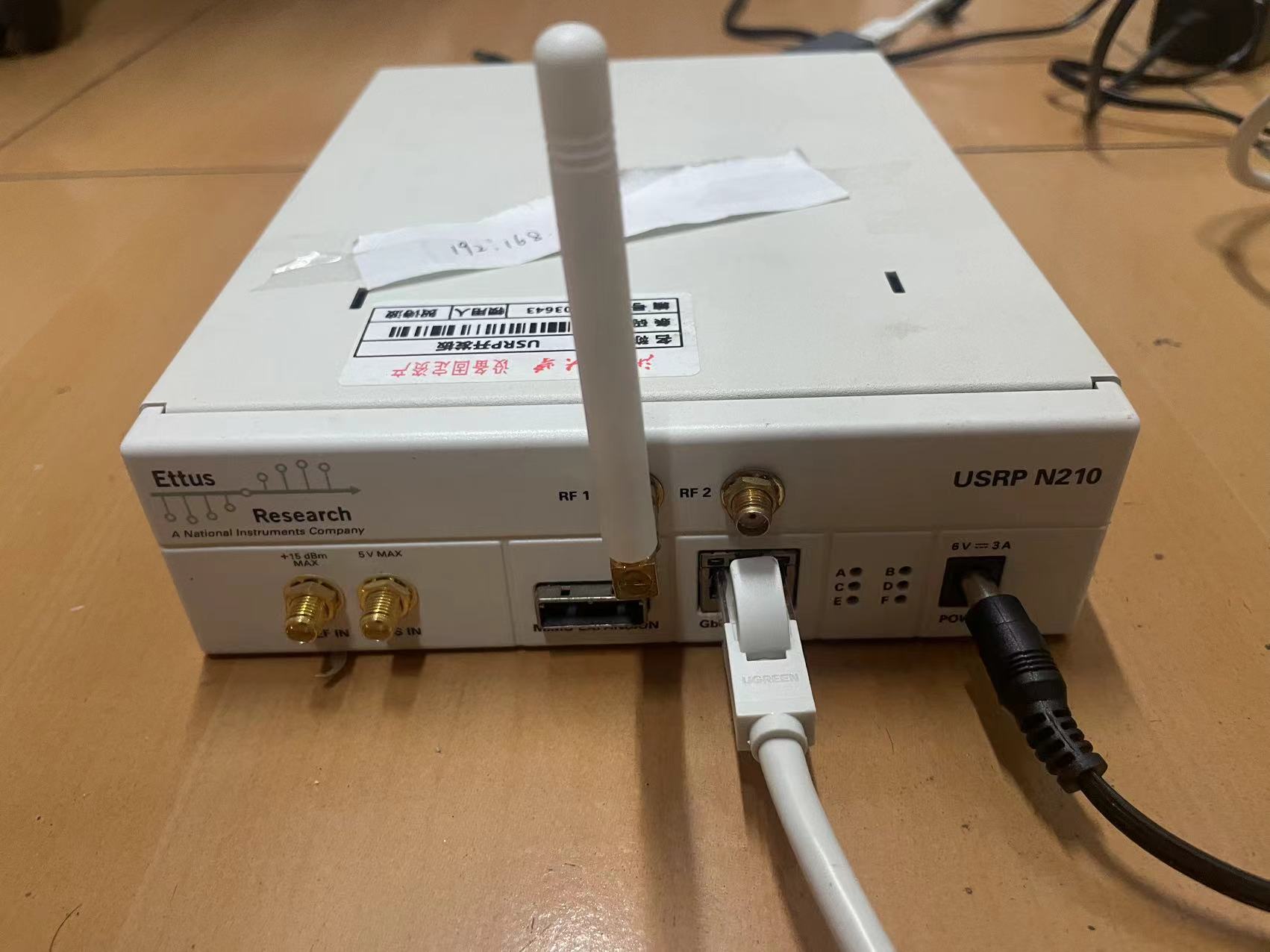}
    \caption{\label{USRP N210}USRP N210.}
  \end{subfigure}
  \begin{subfigure}{.15\textwidth}
    \centering
    \includegraphics[width=\linewidth]{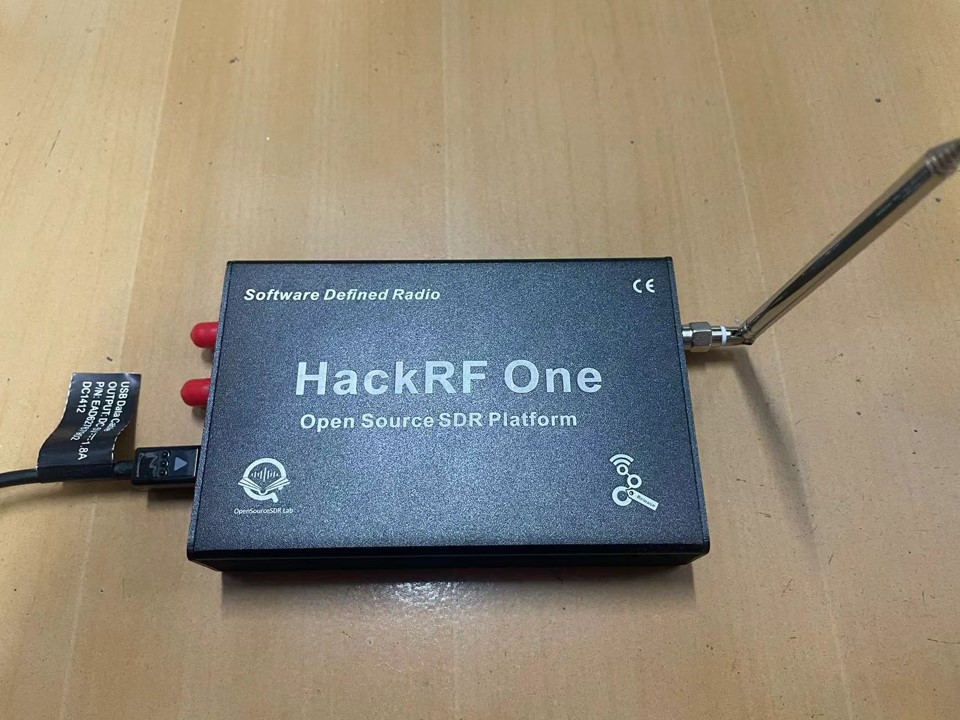}
    \caption{\label{HackRF}HackRF One.}
  \end{subfigure}
  \begin{subfigure}{.15\textwidth}
    \centering
    \includegraphics[width=\linewidth]{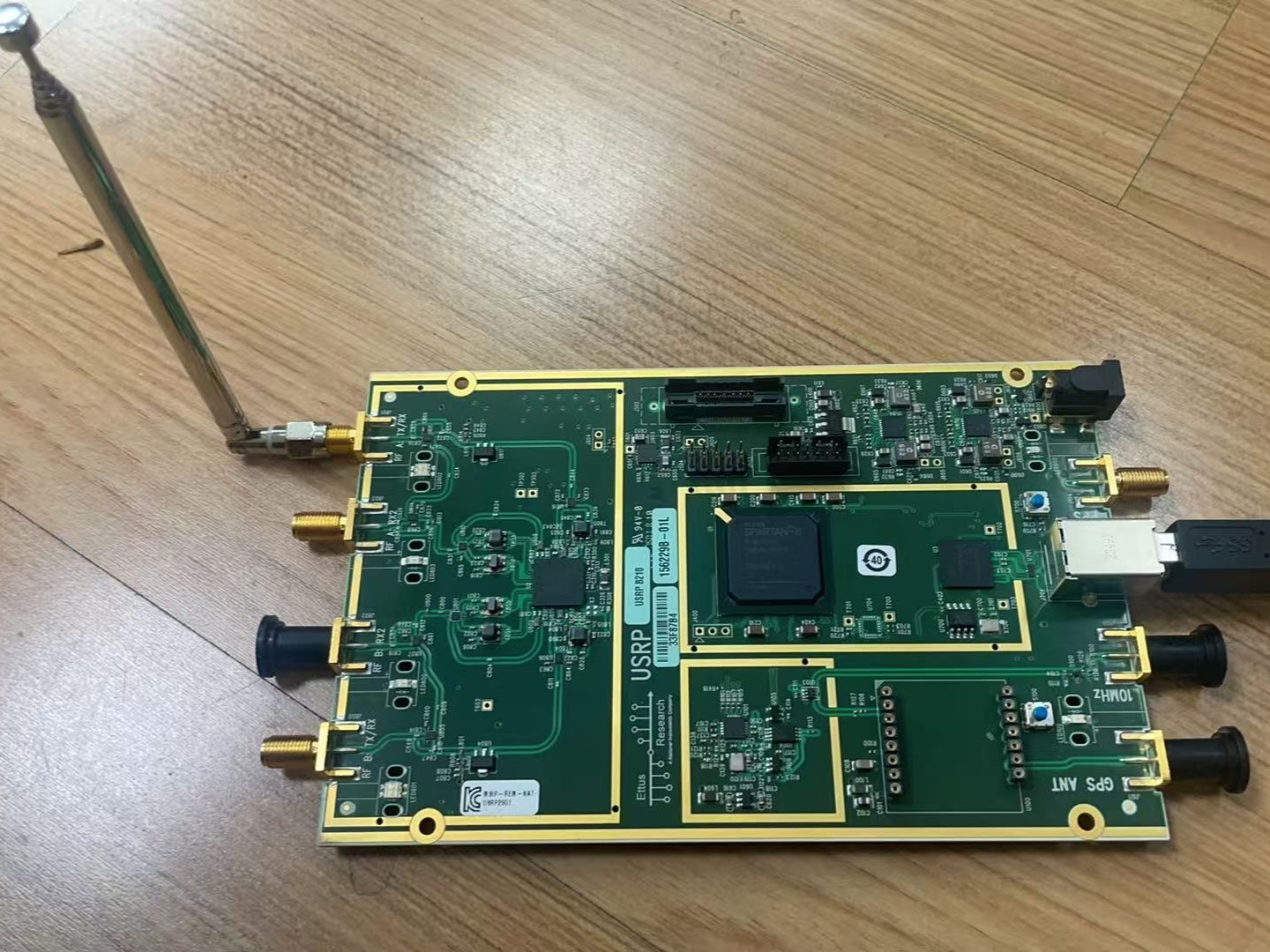}
    \caption{\label{USRP B210}USRP B210.}
  \end{subfigure}
  \caption{Fine tuning and evaluation with different SDR platforms.}
  \label{fig:Implementation}
\end{figure}

The transfer learning framework was validated through cross-device calibration on three SDR platforms: USRP B210, USRP N210 (UBX40 daughterboard), HackRF One. As shown in Fig.~\ref{fig:Implementation}, these three Commercial Off-The-Shelf (COTS) SDR devices manifest varying levels of CFO during practical experimentation, attributable to inherent limitations, including local oscillator instability and constrained frequency division capabilities. To illustrate, when we set the sample frequency as $1.92\,\text{MHz}$ for USRRP N210 equipped with UBX-40 as daughter-board, the return message shows that the actual sample frequency is set as $1923077\,\text{Hz}$. Moreover, according to the datasheet of USRP N210, the local oscillator accuracy is $2.5\,\textit{ppm}$ and the sample rate of Analog-to-Digital Converter (ADC) is $100\,\text{MHz}$, which means the actual sample rate varies from $1.875\,\text{MHz}$ to $1.971\,\text{MHz}$ when it is set as $1.92\times10^6\,\text{Hz}$. These Sampling Frequency Offset (SFO) can introduce severe CFO in signal processing. To establish the dataset of fine tuning, each device collected $1000$ real OFDM frames under indoor multipath conditions. The configuration of the experiment is summarized in Table ~\ref{tab:OFDM cfg}. The intrinsic absence of pilot sequences in our OFDM architecture precludes conventional channel estimation, inevitably introducing phase rotation artifacts at the receiver. To circumvent this impairment, we implement Differential Quadrature Phase-Shift Keying (DQPSK) modulation across subcarriers, leveraging its self-referential phase encoding to eliminate coherent detection dependencies. As illustrated in Fig.~\ref{fig:DQPSK_transmission}, the operational principle of DQPSK resolves the inherent demodulation challenges under phase-ambiguous channels. For the DQPSK system, each symbol's phase $\phi_k$ is determined by:
\begin{equation}
\label{DQPSK expression}
    \phi_k = (\phi_{k-1}+\Delta\phi_k) \mod{2\pi},
\end{equation}
in which $\Delta \phi_k \in \{0,\frac{\pi}{2},\pi,\frac{3\pi}{2}\}$ carries $2$ bits per symbol~\cite{DQPSK}. In this configuration, the transmission of $1,000$ OFDM frames required for device-specific fine-tuning achieves a total airtime of $834$ ms.

\begin{table}[t]
\caption{Configuration of transmitted OFDM Frame.}
\label{tab:OFDM cfg}
\centering
\begin{tabular}{c|c}
\hline
\textbf{Sample Rate (Hz)} & $1.92\times10^6$\\
\hline
\textbf{Symbol Length} & 128\\ 
\hline
\textbf{CP Length} & 32\\
\hline
\textbf{Num of Symbols} & 10\\
\hline
\textbf{Modulation} & DQPSK\\
\hline
\end{tabular}
\end{table}

To mitigate the risk of overfitting during the fine-tuning phase of the DNN, the $1,000$ OFDM frame symbols $S_t$ employed in our fine-tuning dataset were derived from data symbols generated through the modulation of randomly generated bit sequences. This approach ensures the statistical independence and diversity of training samples while maintaining protocol compliance with standard wireless communication frameworks. To assess the feasibility of the receiver-centric architecture, three SDR devices were configured as receivers, while two distinct SDR platforms served as transmitters to emit $1,000$ OFDM frames for system calibration. 
\subsection{Experimental Results}
As the CFO is unobservable in SDR transmission, we cannot accurately assess the accuracy of the estimated CFO. Therefore, we compensate the received signal with the estimated CFO and demodulate it. The constellation of demodulated symbols is shown in Fig.~\ref{fig:Rx_constellation}. It is critically noteworthy that the intentional exclusion of pilot sequences within OFDM frames induces cumulative phase rotation across received symbols. The comparative analysis reveals that while the pre-trained model demonstrates suboptimal performance when directly applied to real-world SDR-received signals, the fine-tuned version (adapted through $1,000$ OFDM frames) achieves significantly enhanced CFO estimation accuracy. As quantified in Fig.~\ref{fig:Rx_constellation}, the refined model outperforms conventional CP-based estimation algorithms under indoor multipath conditions. As QPSK modulation cases the phase of the symbol rather than the amplitude, we perform power normalization for each symbol. Experimental results reveal significant disparities in CFO estimation efficacy across heterogeneous SDR platforms. For high-stability clock architectures exemplified by the USRP N210, conventional CP-based estimation approximately achieves reliable demodulation. In contrast, cost-constrained SDRs like the HackRF One require domain-adapted neural models through fine-tuning to suppress oscillator-induced frequency drift. Notably, pre-trained models demonstrate hardware-agnostic generalization capabilities when applied to USRP B210/N210 platforms without fine-tuning.

\begin{figure}[t]
    \centering
    \includegraphics[width=0.85\linewidth]{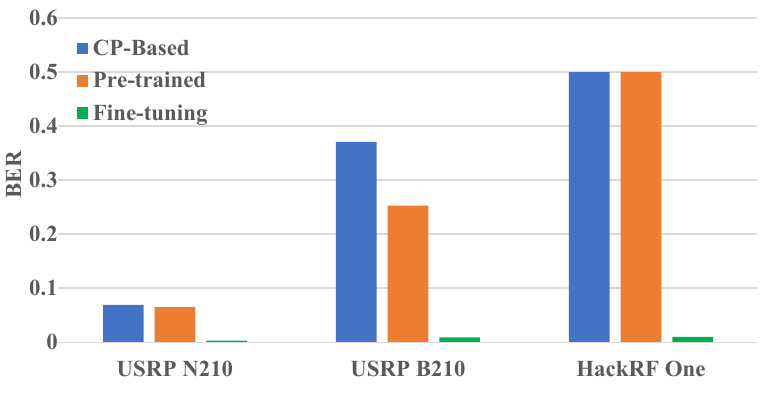}
    \caption{BER analysis across multiple CFO estimation algorithms.}
    \label{BER}
\end{figure}

To systematically evaluate carrier CFO correction efficacy, we conducted a comparative bit error rate (BER) analysis across multiple CFO estimation algorithms under standardized channel conditions. The result is shown in Fig.~\ref{BER}. Experimental results demonstrate that per-device model fine-tuning universally enhances CFO estimation accuracy across the mentioned $3$ SDR platforms. Notably, the HackRF One exhibits the most significant performance gain, where domain-adapted neural estimators reduce bit BER by approximately 48\% (to 1/30 of CP-based levels) compared to non-adapted configurations. 

\section{conclusion}
\label{sec:conclision}
We present a Sim2Real transfer learning framework for device-specific CFO calibration in OFDM systems. By combining hardware-aware simulation pretraining with lightweight receiver adaptation, the method overcomes the limitations of conventional estimators in heterogeneous SDR environments. Key innovations include: \textbf{1)} A parametric distortion model that allows cost-effective DNN pretraining without cross-device data collection; \textbf{2)} Receiver-centric fingerprinting by fine-tuning $1,000$ real frames per device; \textbf{3)} A noise-resistant architecture achieving 48\% BER improvement over CP-based methods. Validated across USRP and HackRF platforms, the approach demonstrates particular effectiveness for low-cost SDRs with inherent oscillator instability. Future work extends to joint CFO-channel estimation and federated calibration strategies.

\bibliographystyle{ieeetr}
\bibliography{reference}
\end{document}